# F-Tree: an algorithm for Clustering Transactional Data Using Frequency Tree

**MAHMOUD A. MAHDI**
Faculty of Computers and Information, Zagazig University

**SAMIR E. ABDELRAHMAN**
Faculty of Computers and Information, Cairo University

**REEM BAHGAT**
Faculty of Computers and Information, Cairo University

**ISMAIL A. ISMAIL**
Faculty of Computers and Information, Zagazig University

## ABSTRACT

Clustering is an important data mining technique that groups similar data records, recently categorical transaction clustering is received more attention. In this research we study the problem of categorical data clustering for transactional data characterized with high dimensionality and large volume. We propose a novel algorithm for clustering transactional data called F-Tree, which is based on the idea of the frequent pattern algorithm FP-tree; the fastest approaches to frequent item set mining. And the simple idea behind the F-Tree is to generate small high pure clusters, and then merge them. That makes it fast, and dynamic in clustering large transactional datasets with high dimensions. We also present a new solution to solve the overlapping problem between clusters, by defining a new criterion function, which is based on the probability of overlapping between weighted items.

Our experimental evaluation on real datasets shows that: Firstly, F-Tree is effective in finding interesting clusters. Secondly, the usage of the tree structure reduces the clustering process time of the large data set with high attributes. Thirdly, the proposed evaluation metric used efficiently to solve the overlapping of transaction items generates a high quality clustering results. Finally, we have concluded that the process of merging pure and small clusters increases the purity of resulted clusters as well as it reduces time of clustering better than the process of generating clusters directly from dataset then refine clusters.

**Categories and Subject Descriptors**

Data mining, Pattern Recognition, Clustering

**General Terms**

Algorithms

## 1. INTRODUCTION

Transactional data is a kind of special categorical data where records are made up of non-numerical attributes. Transactional data are generated by many applications such as e-commerce, healthcare, CRM, and so forth [1]. Transaction data play an important role in many fields like Market basket data, Web usage data, Customer profiles, Patient symptoms records, and image features [2]. A transactional data set consists of $N$ transactions, each of which consists of varying number of item $i$. The size of transactional data is usually large, so there is great demand for fast and high quality algorithms for clustering large scale transaction datasets.

In general, the main goal of clustering transactional data set is to generate clusters that have the items of the same occurrence, so the transactions in clusters are similar; in other words, they maximized the occurrence of items in each cluster. To do that, different approaches have been proposed as shown in the next section.

There are different approaches for clustering data. They fork from two main types of clustering techniques, hierarchical clustering and non-hierarchical clustering. Hierarchical clustering produces a nested series of partitions while non-hierarchical produces only one [3]. But non-hierarchical divides the point space into K clusters that optimize a certain criterion function [4]. The hierarchical clustering algorithm forms clusters in hierarchical fashion [5] so, the number of clusters at each step is fewer than the previous one. Hierarchical clustering is divided in two different approaches:

1. Top-down divisive approach (Descendant, or Document).
2. Bottom-up agglomerative approach (Ascendant).

The first approach starts with big cluster and recursively split each cluster if advantageous. It provides hierarchical trees where terminal segments represent a partition of variables in the same cluster. The other approach starts with one cluster per point, then iteratively find two clusters to merge if advantageous, finally clusters are formed by finding pair with maximum similarity. This method leads to a hierarchy of rested clusters and based on the choice of a similarity coefficient and aggregation strategy [6].

On the other hand, we proposed a new technique to fast create small clusters with same items occurrence. This can speed efficiently handling large transactional datasets. That is our strategy; it is based on document hierarchical clustering strategies in dropping the number of cluster down. The difference between the document hierarchical and our proposed approach is that we do not start with one large cluster then recursively split it down many times to reach the best clusters number. But we generate small clusters; then we merge most similar clusters together depending on the proposed criterion function that tries to increase the intra-cluster overlapping of transaction items by increasing the probability of large item of the cluster.

Our approach has 4 unique features:

1. We present a new clustering technique based on the frequent tree structure, named F-Tree.
2. We present a new concept for a categorical similarity measurement based on the criterion probabilities of overlapping to efficient clustering of transactional datasets, which make our implementation more scalable and dynamic.
3. We develop the overlapping estimator technique to estimate the overlapping degree between the clusters, so we do not need any input parameter from user to cluster or refine the data.
4. We implement the F-Tree clustering algorithm and the proposed metric using a full automated and scalable transactional clustering framework, called FCSO. The FCSO framework is designed to combine the F-Tree algorithm with the overlapping estimator that can automatically tune the degree of overlapping, which is the important parameter of clustering algorithm.

Experiment shows that our algorithm runs much faster than LargeItem [7] algorithm with high clustering quality than CLOP [2], ROCK [4] and Seed [8] algorithms.

## 2. RELATED WORK

In this section we look at four different algorithms in clustering transaction. Section 2.1 discuses the LargeItem that is based on the large items in the transaction dataset. Section 2.2 discusses the Small Large Ratio SLR that is speeding-up the LargeItem approach. Section 2.3 discusses a new clustering algorithm based on the seed clusters. And section 2.4 discusses a clustering algorithm based on the concept of large items and coverage density.

### 2.1. Clustering Transactions Using Large Items

LargeItem algorithm uses the concept of large items to cluster transactions [7]. Their approach measures the similarity of a cluster based on the large items in the transaction dataset. Without using any measure of pair-wise similarity the large item algorithm group's categorical data by iterative optimization of global criterion function. The criterion function is based on the large item; an item is consider large in a cluster of transaction if it have occurrence rates larger than a minimum support parameter specified by the user [7].

The large item approach scanning each transaction and either allocated it to an existing cluster or assigned to a new cluster based on a cost function. The cost function measures the degree of similarity between a transaction and a cluster based on the number of large and small items shared between that transaction and the given cluster.

LargeItem method is similar to k-means algorithms in that it scans transactions and assigns the next transaction to the "best" cluster. But, two differences exist. First, LargeItem approach does not require the number $k$ of clusters. K-means algorithms, on the other hand, assume a fixed number $k$ of clusters, thus, cannot be applied to applications where the number of clusters can evolve. Second, choosing a cluster for the next transaction not based on the distance between the cluster and the transaction, but based on the global goodness of clustering. This goodness measures by minimizing the total cost. The total cost is determined by equation (1).

$$Cost(C) = w \times Intra(C) + Inter(C) \qquad (1)$$

Where $Intra(C)$ measures the total number of small items which represent the item-dissimilarity, as defined by equation (2).

$$Intra(C) = |U_{i=1}^{k} small_i| \qquad (2)$$

Where $k$ is the current number of clusters. But $Inter(C)$ measures the duplication of large items in different clusters, where a large item is an item whose support exceeds the minimum support $\theta$, which represent the item similarity. $Inter(C)$ is defined by equation (3).

$$Inter(C) = \sum_{i=1}^{k} |large_i| - |U_{i=1}^{k} large_i| \qquad (3)$$

If the weight $w > 1$; then $Intra(C)$ will be more important than $Inter(C)$, and vice versa. LargeItem needs to set the support $\theta$ and the weight $w$.

The LargeItem algorithm included two phases. The Allocation phase and the Refinement phase. In allocation phase, the database is scanned once and each transaction $t$ is read in sequence, each $t$ can be assigned to an existing cluster or new cluster is created for $t$, to minimize total $Cost(C)$ for the current clustering $C$. The cluster identifier of each transaction is written back to the file. In Refinement phase, each transaction $t$ is read and moves $t$ to an existing cluster to minimize $Cost(C)$ and may stay where it is. After each move, the cluster identifier is updated and any empty cluster is eliminated immediately. When no transaction is moved in one pass of all transactions. Refinement phase terminates;

otherwise, a new pass begins. In each step the criterion $Cost(C)$ is optimized. The key step is finding the destination cluster for allocating or moving transaction $t$ [7].

In general the LargeItem is exhaustive in the decision procedure of moving a transaction $t$ to the best cluster.

## 2.2 An Efficient Clustering Algorithm Based on Small Large Ratio

To speed-up the LargeItem method, a new method called Small-Large Ratio SLR [9] was introduced. This method basically uses the measurement of the ratio between small to large items and utilize this ratio to perform the clustering on the transaction. The ratio of the number of small items to that of large items in a group is called small-large ratio (SLR) of that group. With one attribute $I$, $|L_I(t)|$ represents the number of the large items in $t$ and $|S_I(t)|$ represents the number of the small items in $t$. The SL ratio of $t$ with attribute $I$ in cluster $C_i$ is defined by equation (4).

$$SLR_I(C_i, t) = \frac{|S_I(t)|}{|L_I(t)|} \qquad (4)$$

The main goal of algorithm is to minimize the SLR in each group. The procedure of this algorithm includes also two phases, the allocation phase and the refinement phase. The method of allocation phase is straightforward and the approach taken in LargeItem will be enough. The use of the same function to compute the total cost when allocating a new transaction but exceed the maximal ceiling $E$ (the minimum number to appear an item in transactions). When counting the small items. In the refinement phase, each transaction will be evaluated for its status to minimize the total cost. The goal of this method focuses on designing an efficient algorithm for the refinement phase.

The improvement of this method is a result of inefficient steps in the refinement phase of the LargeItem algorithm. This could be partly due to the reason that the similarity measurement used in LargeItem approach does not take into consideration the existence of small items. The SLR solves this problem by proposing a maximal ceiling $E$ for identifying the items of rare occurrences. If an item whose support is below a specified maximal ceiling $E$, that item is called a small item. So, small items in a cluster contribute to dissimilarity in a cluster, algorithm SLR can efficiently determine the next cluster for each transaction in iteration of refinement procedure. The SLR algorithm compares the SL ratios with the pre-specified SLR threshold $\alpha$ to determine the best cluster for each transaction.

In general this algorithm must compute all costs of new clustering when transaction $t$ is put into another cluster, by utilizing the concept of small-large ratios.

## 2.3 Transaction Clustering Using a Seeds Based Approach

A new approach to solve the problem of transaction clustering based on an initial seeding of cluster centroids was proposed [8]. The [8] concluded that both the large item [7] and SLR [9] method suffers a common drawback; that they may fail to give a good representation of the clusters.

Overall the clustering approach is divided into two main phases: seed generation and allocation phases. The algorithm starts the method by finding the optimal number of clusters. The initial choosing of seeds are the large items in the dataset and this begin by setting a minimum support threshold. For a large item set to be considered a cluster seed the frequency of co-occurrence of all pairs of subsets within the seed must occur together with a frequency above a threshold value at a given significance level. This effectively ensures that cluster seeds of size ≥ 2 have items that co-occur together at a frequency that is statistically significant. In addition, it requires that all cluster seeds satisfy an improvement constraint when they are extended.

The seeds produced in the initial phase are considered as the initial centroids for the clusters, In the Allocation Phase, transactions are assigned to clusters on the basis of similarity to cluster centroid. In order to measure similarity a modified version of the Jaccard similarity coefficient [10] used for each transaction $t$ as equation (5).

$$sim(t, c_k) = \frac{|t \cap c_k|}{|t \cup c_k| - |t \cap c_k| + 1} \qquad (5)$$

The algorithm calculates the similarity between $t$ and the existing centroid $c_k$. Allocated all transactions to clusters, the refine is done by re-computing the centroids belonging to transactions allocated to a given cluster. The updating of centroids will result in the need for reorganization of the clusters, this process of centroid update and cluster reorganization will be repeated until a suitable point of stabilization of fitness function is reached. The fitness measure calculates the average similarity between every transaction in a cluster to its centroid. This fitness function is defined by equation (6).

$$J = \frac{1}{k} \sum_{j=1}^{k} \frac{\sum_{t \in c_j} d(t, c_j)}{|C_j|} \qquad (6)$$

This approach tried to maximize this function value.

## 2.4 Efficiently Clustering Transactional Data with Weighted Coverage Density WCD

Weighted coverage Density WCD algorithm introduces a new concept beside the clustering algorithm. The [1] developed two evaluations measurement based on concept of large items and coverage density respectively, First evaluation measurement is large item size ratio

(LISR) that uses the percentage of large items in the clustering result to evaluate the clustering quality Second evaluation measurement is average pair clusters merging index (AMI) applies coverage density to indicate the structured different between clusters. The key design idea of the WCD algorithm is the definition of the weighted coverage density based clustering criterion [1]. This approach tries to maximize the frequent items as possible with clusters and make controls items overlapping between clusters. The weighted coverage density of a cluster $c_k$ is defined by equation (7).

$$WCD(C_k) = \frac{\sum_{j=1}^{M_k} occur(I_{Kj})^2}{S_{k*N_k}} \qquad (7)$$

Where $M_k$ is the number of distinct items in a cluster $c_k$, $I_k$ is the item set $I_k = \{I_{k1}, I_{k2}, \ldots, I_{kM_k}\}$, and $N_K$ is the number of transaction in a cluster and $S_k$ is the sum of occurrences of all items in cluster $C_k$. They also defined the clustering criterion function as expected weighted coverage density (EWCD) by equation (8).

$$EWCD = \frac{1}{n} \cdot \sum_{k=1}^{k} \frac{\sum_{j=1}^{mk} occur(I_{Kj})^2}{S_k} \qquad (8)$$

The algorithm is based on EWCD function and tries to maximize the EWCD criterion. By default when all transaction is considered in single cluster it will get the maximum EWCD, since this function cannot determine when algorithm have to stop because merging clusters maximize the EWCD; so an additional phase priori to clustering phases is required to determine the best number of clusters by take a sample of data and run at different value of $K$. But that make algorithm poor in dynamic environment as number of clusters can be changed suddenly.

## 3. TRANSACTION CLUSTERING USING FREQUENT TREE

**Notations** Throughout this research we use the following notations. A transactional dataset $D$ is a set of transaction $\{t_1, \ldots, t_n\}$. Each transaction $T$ is a set of items $\{i_1, \ldots, i_m\}$. A clustering $\{C_1, \ldots, C_k\}$ is a partition of $\{t_1, \ldots, t_n\}$, that is, $C_1 \cup C_2 \cup C_3 \ldots \cup C_k = \{t_1, \ldots, t_n\}$ and $C_i = \emptyset \wedge C_i \cap C_j = \emptyset$ for any $1 \leq i, j \leq k$. Each $C_i$ is called a cluster, $n$, $m$, and $k$ are used respectively for number of transactions, number of items, and number of clusters. The frequent support of the item in the dataset is represents by $|I_i|$.

### 3.1. Introduction

Unlike traditional data clustering, transaction clustering requires transaction to be partitioned across clusters in such a manner that instances within a cluster share a common set of large items, where the concept of the large follows the same meaning attributed to frequent items in association rule mining [11]. Thus it is clear that transaction clustering requires a fundamentally different approach from the traditional clustering technique. So that the term Frequent Tree "F-Tree" comes from the Frequent Pattern Tree "FP-Tree" algorithm that works on mining association rules.

In this section we will present a new approach for transaction clustering that is based on small of initial clusters. The approach consists of two phases: An initial clusters generation in the allocation phase followed by a clusters merging in the refinement phase. In the allocation phase the clusters are identified from F-Tree approach at cut level $l$. Once initial clusters are generated, the next phase merges most similar clusters together. In order to group most similar cluster, we present our new fitness function to solve the overlapping problem between clusters as we will discuss later.

**3.2. Allocation phase**

In this implementation the allocation phase is done by three main steps. The first is the preprocessing step which is used to determine the frequencies $f_i$ of all items in the dataset. The second step is to build the F-Tree structure, and The Finial step is to extract the initial clusters by pruning the F-Tree.

**3.2.1. Preprocessing**

The F-Tree requires scan initially the dataset $D$ to determine the frequencies $f_i$ of the items $I_i$ (the support of single element item sets). In addition, items in each transaction are sorted, such that they are in descending order with respect to their frequencies in the dataset.

*Example 1*: To demonstrate the algorithm we will explain the sample of transaction dataset in Table 1 throughout this paper, which shows an example of a simple transaction dataset. First the preprocessing is to get the frequencies of the items in this dataset, sort descending, and the result of this is shown in Table 2 using the dataset in Table 1.

Table 1. Sample of Transaction dataset

| TID | Items |
|---|---|
| 1 | {A,B,F} |
| 2 | {B,C,D} |
| 3 | {A,C,D,E} |
| 4 | {A,D,E} |
| 5 | {A,B,C} |
| 6 | {A,B,C,D} |
| 7 | {B,C,F} |
| 8 | {A,B,E} |
| 9 | {A,B,D} |
| 10 | {B,C,E} |

Table 2. Items Frequencies of Sample on Table 1

| Item | Support |
|---|---|
| B | 8 |
| A | 7 |
| C | 6 |
| D | 5 |
| E | 4 |
| F | 2 |

### 3.2.2. Frequent Tree Concept

F-Tree approach models a categorical dataset as a $l$ level tree; where each node in the tree corresponds to one item value with it frequency, and the item at level $k$ has a higher frequency than its children at the deeper level $l + 1$. The groups of items that get in the path starting from the tree root to any leaf node composing a single transaction, so all paths from root to leaves nodes compose all transactions in the dataset. Thus, each path represents a set of transactions that share the same prefix.

### 3.2.3. Building the F-Tree

In our implementation the F-Tree is built with a straightforward procedure: (1) reading a transaction $t_i$ from a dataset $D$, (2) order the list of items in the transaction $t_n = \{I_1, ..., I_m\}$ by their frequencies $f_i$ where $|I_i| > |I_{i+1}|$, and (3) inserting the items in the sorted list of transaction into an initially empty F-Tree by starting from the root of tree. Inserting item $I_1$ in the node to be child of the root with $f_1 = 1$ then insert item $I_i$ on the node to be child of the child of the node of item $I_{i+1}$ with $f_i = 1$ and so on, at the leaf node we are holding the transaction number $t_i$. The remaining transaction is inserted in the same way but if a node is found holding the same item value $I_i$ and referring to the same parent node at level $l - 1$; we increase its frequency $f_i = f_i + 1$ without inserting a new node.

At first sight, it may seem to be normal to build F-Tree by inserting transaction after transaction and creating the necessary nodes for each new transaction. In fact, such an approach has the advantage that the transaction dataset needs not to be loaded into main memory. Since only one transaction is processed at a time, only the F-Tree representation and one new transaction are in main memory. This typically saves memory space and avoids costly datasets scans, because an F-Tree is much more compact representation of a transaction dataset.

In this Implementation, an F-Tree node contains data structure fields for (1) an item identifier, (2) a counter, (3) a pointer to the parent node, (4) a list of pointer to successor node, and (5) a transaction identifier used only in leaf nodes. This operation is demonstrated in Figure 1, which shows an F-Tree of the transaction dataset found in Table 1 and using result form Table 2.

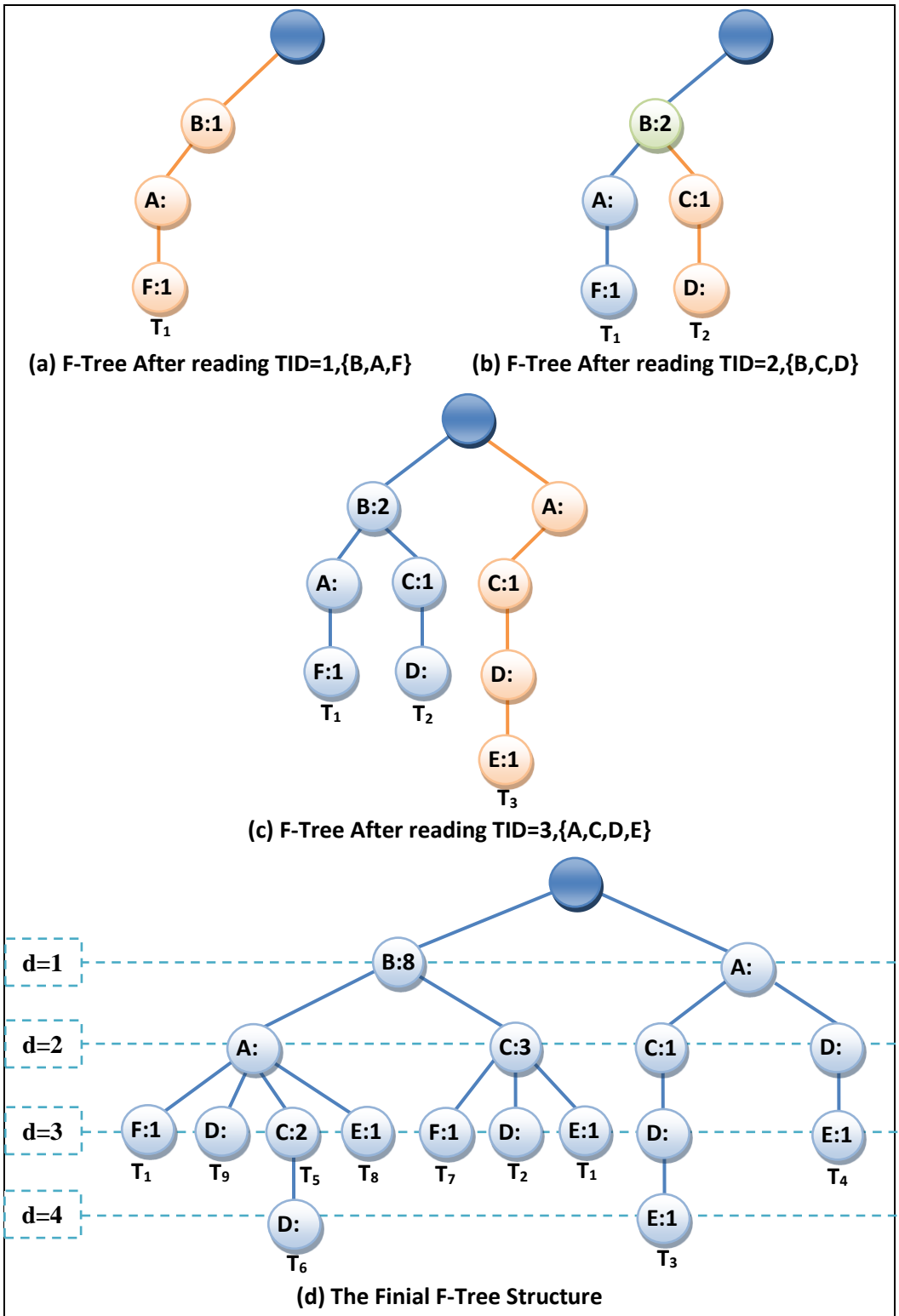

**Figure 1. The Building F-Tree Structure for Transactions in Table 1**
Note. Items are sorted in decreasing support count before inserted into the tree

### 3.2.4. Extracting the Initial clusters

To extract the clusters from the F-Tree we first prune the F-Tree at level $l$, as each node at each level group the same items from different transactions. We make the pruning level equal to the minimum support of clustering process with respect to the maximum depth of the F-Tree using the equation (9),

$$pruning\ level = \min support\ \times \max depth \quad (9)$$

The $\min support\ \theta$ must be large to generate high purity cluster, because once item is assign to cluster there is no chance to remove it from that cluster as our method to cluster transactional data is to generate high pure clusters to ensure that merging will also lead to high pure clusters.

To illustrate how the clusters are extracted, we continue from F-Tree Structure from Figure 1. Let the $\min support\ \theta = 100\%$, and the $\max depth = 4$ then the $pruning\ level = 100\% \times 4 = 4$. The extracted clusters result shown in Table 1. Table 4 and Table 5 show the extracted clusters from the $pruning\ level = 2$, and the $pruning\ level = 3$ respectively, While Table 6 show the extracted clusters from $pruning\ level = 1$. Normally when pruning the F-Tree at the $level = 1$, the number of clusters equal to the number of transactions if there is no repeating in the transactions.

Table 3. The Clusters Extracted form Level=4

| Cluster | Transaction |
|---|---|
| 1 | $\{T_1, T_2, T_5, T_6, T_7, T_8, T_9, T_{10}\}$ |
| 2 | $\{T_3, T_4\}$ |

Table 4. The Clusters Extracted form Level=3

| Cluster | Transaction |
|---|---|
| 1 | $\{T_1, T_5, T_6, T_8, T_9\}$ |
| 2 | $\{T_2, T_7, T_{10}\}$ |
| 3 | $\{T_4\}$ |
| 4 | $\{T_3\}$ |

Table 5. The Clusters Extracted form Level=2

| Cluster | Transaction |
|---|---|
| 1 | $\{T_1\}$ |
| 2 | $\{T_5, T_6\}$ |
| 3 | $\{T_8\}$ |
| 4 | $\{T_9\}$ |
| 5 | $\{T_2\}$ |
| 6 | $\{T_7\}$ |
| 7 | $\{T_{10}\}$ |
| 8 | $\{T_4\}$ |
| 9 | $\{T_3\}$ |

Table 6. The Clusters Extracted form Level=1

| Cluster | Transaction |
|---|---|
| 1 | $\{T_1\}$ |
| 2 | $\{T_5\}$ |
| 3 | $\{T_6\}$ |
| 4 | $\{T_8\}$ |
| 5 | $\{T_9\}$ |
| 6 | $\{T_2\}$ |
| 7 | $\{T_7\}$ |
| 8 | $\{T_{10}\}$ |
| 9 | $\{T_4\}$ |
| 10 | $\{T_3\}$ |

### 3.2.5. F-Tree Algorithm

The major steps of F-Tree approach as we discussed are: (1) Computing the items support. (2) Building the F-Tree. And (3) Extracting the Clusters. Figure 2 shows the overview of allocation phase the first part of our algorithm.

```
/* Allocation phase */
Input: Transaction dataset, D = {t_1, ..., t_n}, minimum support θ value
Output: Initial clusters, S = {s_1, ..., s_k}

     // Scan the dataset to determine the items support
(1)  for each transaction T in the dataset D
(2)    for each item I_i in the transaction T
(3)      FrequentList[I_i] + +

     // Building the F-Tree
(4)  for each transaction T_ID in the dataset D
(5)    sort items I_i in transaction T_ID by FrequentList
       where I_n = {I_1, ..., I_n} and |I_i| ≥ |I_{i+1}|.
(6)    CurrentNode ← root
(7)    for (i = 1 to n).
(8)      if CurrentNode does not contain ChildNode(I_i)
(9)        CurrentNode. Add ChildNode(item = I_i, frequence = 1)
(10)     else
(11)       CurrentNode. ChildrenNode(I_i). frequence + +
(12)     CurrentNode ← CurrentNode. ChildrenNode(I_i)
(13)   CurrentNode. TransactionList. Add(T_ID)

     //Extracted Initial Clusters
(14) for each node(N_k) at level = min support × max depth
(15)   Create new Cluster C_j
(16)   for each Transaciton T_ID in the path rooted by node (N_k)
(17)     Cluster(C_j). TransactionList. Add(T_ID)
(18)   for each Item I_i in the path from root via the node(N_k) to leaves
(19)     Cluster(C_j). Items. Add(I_i)

(20) Return Clusters List
```

**Figure 2. The allocation phase algorithm of F-Tree clustering**

### 3.3. Refinement phase

The allocation phase generates small initial clusters that will be used in the refinement phase to generate a final cluster. The main goal of refinement phase is to merge the similar clusters together to get the final view of clusters and this process accomplished by applying our criterion function to measure the overlapping degree between clusters. Naturally, most algorithms depend on a criterion function that is used to measure the goodness of clusters. The criterion function can be defined locally or globally.

Approaches based on local function are done by computing evaluation function between each item inside the cluster itself; the result shows the degree of how items inside a cluster are related to each other. On the other side, approaches based on global are done by computing evaluation function between clusters; the result shows the degree of how clusters are dissimilar and more distinct. However for large datasets, the computational cost of these local approaches is heavy compared with global approaches and that was our goal in our implementation and in choosing the criterion function. Although the related work used either a criterion function that is composed of local and global approaches together as [7], and [9], or two criterion functions one for each as in [1], [11]. But in our clustering technique, we use only global method to measure the goodness of the cluster and to make our technique very fast. The key of method here is we only compute the similarities between clusters not between transactions; that is because clusters are fewer than transaction, what make this method faster enough than others methods.

In our implementation the refinement phase groups the most similar clusters resulting from the allocation phase. The grouping process is accomplished by the fitness function that is based on the probabilities of intra-join items which tries to keep as many frequent items as possible within clusters and controls the items overlapping between the clusters implicitly.

### 3.3.1. The intra-join of weighted items concept

In this section we introduce a new measurement function that estimate the goodness of clusters. We use a similarity measure function in order to determine the best pair of clusters to merge at each step of the refinement phase.

The key difference between most methods is in defining criterion function of evaluation clustering, but the difficulty was in proposing a good scenario to solve the overlapping between clusters which change according to the behavior of the dataset. So, to solve the overlapping problem we need to evaluate two opposite requirements. The first is to maximize the frequent items within clusters, and the second is to minimize the items overlapping between clusters.

Using the probability estimation for overlapping computation has many advantages. Probabilities are a good evaluation for further processing and intuitive. Given a cluster $c_k$, suppose the number of distinct items is $M_k$, the items set of $C_k$ is $I_K= \{I_{k1}, I_{k2}, \dots, I_{kM_k}\}$, the number of transactions in the cluster is $N_k$, and the sum of occurrences of all items in cluster $C_k$ is $S_k$ and given by equation (10).

$$S_k = \sum_{j=1}^{M_k} |I_{kj}| \qquad (10)$$

Now, the weight of item $W_j$ inside a cluster $C_k$ is defined as the ratio of occurrences of an item $J$ to the sum of occurrences of all items inside the cluster; in other words the probability of an item inside the cluster $C_k$, shown by equation (11).

$$W_j = P(I_j) = \frac{|I_j|}{S_k} = \frac{|I_j|}{\sum_{j=1}^{M_k}|I_{kj}|} \qquad (11)$$

$$\text{Here } \sum_{j=1}^{M_k} W_j = 1$$

We present the definition of the probability of the overlapping between cluster $C_i$ and $C_j$ by estimating intersect items between clusters, as defined by equation (12).

$$sim(C_i, C_j) = \sum_{k=1}^{|C_i \cap C_j|} W_{k \epsilon C_i} \times \sum_{k=1}^{|C_i \cap C_j|} W_{k \epsilon C_j} \qquad (12)$$

### 3.3.2. The Neighbor Threshold

A cluster's neighbors are those clusters that are considerably similar to it. Let $sim(C_i, C_j)$ be a similarity function that normalizes and captures the closeness between the pair of clusters $C_i$ and $C_j$. We conclude that $sim$ values are between **0** and **1**, with larger values indicating that the clusters are more similar. Given a threshold $\alpha$ between **0** and **1**, a pair of cluster $C_i, C_j$ are defined to be *neighbors* if equation (13) holds

$$sim(C_i, C_j) \geq \alpha \qquad (13)$$

In equation (13), $\alpha$ is a parameter that can be used to control how close a pair of clusters must be in order to be considered neighbors. Thus, higher values of $\alpha$ correspond to a higher threshold for the similarity between a pair of points before they are considered neighbors. Assuming that $sim$ is **1** for matching clusters and **0** for totally dissimilar clusters, a value of 1 for $\alpha$ constrains a cluster to be a neighbor to only other identical clusters. On the other hand, a value of **0** for $\alpha$ permits any arbitrary pair of clusters to be neighbors.

### 3.3.3 Merging algorithm

Here we call the refinement phase the merging phase. The major steps of the merging algorithm are: (1) compute the similarity list between clusters. (2) Creating the group of similar clusters. (3) Merging clusters in the same group. Figure 3 shows the overview of refinement phase the second part of our algorithm.

To compute the similarity list we recursively try to find the maximum similar cluster for all clusters, whose values are are also larger than the merging overlap threshold $\alpha$, Table 7 displays the first similarity list for clusters shown in Table 3

Table 7. The Maximum Similarity list between Clusters

| Cluster $C_i$ | Cluster $C_j$ | Max Sim($C_i$,$C_j$) | Transaction in $C_i$ |
|---|---|---|---|
| 1 | 2 | 0.96 | {$T_1$,$T_5$,$T_6$.$T_8$,$T_9$} |
| 2 | 1 | 0.96 | {$T_2,T_7,T_{10}$} |
| 3 | 4 | 0.81 | {$T_4$} |
| 4 | 3 | 0.81 | {$T_3$} |

If the overlap threshold $\alpha = 0.8$. Then the group of similar cluster can be shown in Table 8. If we merge every cluster resulting Table 8 into it similar group, we get the cluster with the transaction list as shown in Table 9.

Table 8. the similar cluster groups

| $G_i$ | Cluster $C_j$ |
|---|---|
| 1 | 1,2 |
| 2 | 3,4 |

Table 9. The similar cluster transactions

| Cluster $C_i$ | Transaction in $C_i$ |
|---|---|
| 1 | {$T_1,T_5,T_6.T_8,T_9,T_2,T_7,T_{10}$} |
| 2 | {$T_3, T_4$} |

```
/* Refinement phase */
Input: Initial cluster list, S = {s_1, ..., s_k}, Merging overlap threshold, α
Output: Clusters List, C = {C_1, ..., C_m}

      // Starting by the Initial cluster
(1)   C ← S
(2)   do
(3)      m = C.Count

         // Compute the Similarity List
(4)      SimList ← ∅
(5)      for each cluster C_i in the Cluster List C
(6)         [ SimList[i] ← j {max{j → Sim(C_i, C_j)|Sim(C_i, C_j) > α}, C_j ∈ C}

         // Create Group for Similar Cluster G
(7)      G ← ∅
(8)      for each cluster c_l in the SimList[l]
(9)         [ G_l ← {l ∪ SimList[l], l ∈ G_l, OR SimList[l] ∈ G_l}

         // Merge Similar Group G
(10)     C ← ∅
(11)     for each group G_l in G
(12)        for each cluster c_h in group G_l
(13)           for each transaction t in cluster c_h
(14)              [ C ← {C_n ∪ t}

         // Repeat (4) to (12) Until no further goodness merging
(15)  Until ( m = C.Count )

(16)  Return Clusters C
```

Figure 3. The Refinement phase algorithm of F-Tree clustering

## 4. IMPLEMENTING F-TREE ALGORITHM WITHIN FCSO FRAMEWORK.

In this section, we present our idea in automate the clustering technique, and describe the implementation of the fully automated framework, named FCSO (F-Tree Clustering using Sample Overlap).

### 4.1. The Overlap Estimator

The closeness of clusters changes according to the dataset; as the overlapping between clusters varying depends on the behavior of the dataset. We propose an automated framework to specify the best value of neighbor's similar clusters according to the training dataset. The framework tries to find the minimum value of closeness of neighbor's cluster $\alpha$ that will result 100% purity of clusters, and use this value in clustering the transaction of dataset. The algorithm starts the test by computing purity with setting $\alpha$ from **1** to **0** stepping **0.1**, and then detecting the minimum value of $\alpha$ that gives the purity of 100% to be the overlap threshold value.

### 4.2. The Implementation of the FCSO Framework

The FCSO framework is designed to perform the transactional data clustering in four steps, the first step is to estimate the overlap threshold value as we presented in section 4.1. The others steps are of the allocation phase and the refinement phase of F-Tree algorithm. In additional merging the similar clusters is based on the overlap threshold value that returned from the overlap estimation step. Figure 4 depicts the overall process flow-diagram of our proposed system.

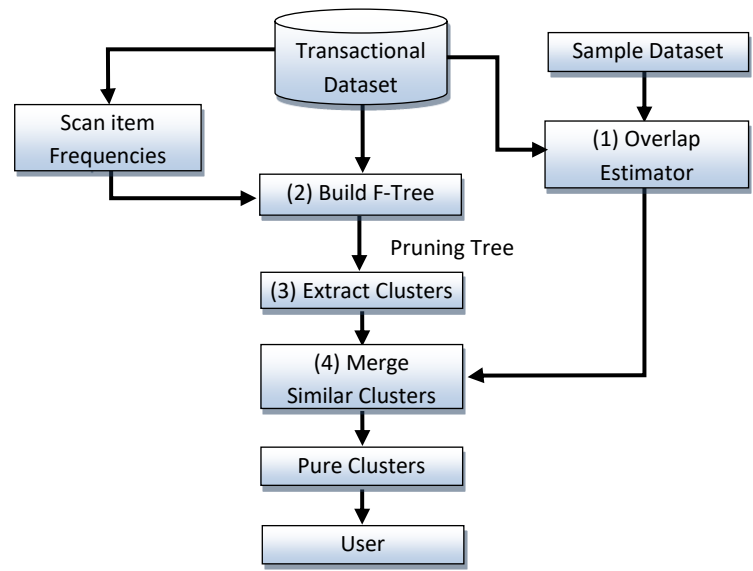

**Figure 4. Flow diagram of F-Tree Clustering process and overlap estimator computation**

## 5. EXPERIMENTS RESULTS

In this section, we analyze the accuracy, and the execution time of F-Tree with two real-life datasets. We also conducted several experiments for clustering to assess the general performance of the F-Tree algorithm.

All experiments have been performed on a Core2 due 2.33Mhz with 2.5 GB of main memory, with a running operating system Windows XP professional service pack 3.

### 5.1. Datasets

Our experiments have used two real datasets: Zoo and Mushroom from the UCI machine learning repository [12].

**Zoo**. Is a real dataset; it contains 101 data records for animals. Each data record has 18 categorical attributes (animal name ,15 Boolean attributes ,1 numeric with set of values [0,2,4,5,6,8] and animal type values 1 to 7 to describe the features of animals. the animal name and animal type values are ignored in our transformed file, while the animal type also serves as an indication of domain clustering structure.

**Mushroom**. Is a real dataset, it contains 8124 instances, which is also used for quality testing Each data record has 22 categorical attributes (e.g. cap-shape, cap-color ,habitat etc.) and is labeled either "*edible*" or "*poisonous*". The dataset contains 23 species of mushroom according to the literature.

### 5.2 Performance Evaluations

The first stage of analysis involves an overall comparison of algorithms execution time. Secondly we make a deep analysis for the cluster quality. We compare the execution speed of F-Tree on Mushroom with LargeItem [7]. And we try our LargeItem implementation to get the direct result.

We run the two algorithms with different minimum support. The results of this test are shown on Figure 5. The F-Tree algorithm uses the merge probability $\alpha = 0.8$, while the LargeItem use the weight of intra $w = 1$. From experiment result we noted that the minimum execution time of LargeItem algorithm could be obtained when use we the weight $w$ of intra equal to 1, and as the weight of the intra increases the execution time increased. In contrast; the maximum execution time of F-Tree algorithm could be obtained when the merge probability equal to 100% and the execution time is decreased as overlap threshold value decreased.

Form Figure 5, we can conclude the large different in execution time between LargeItem algorithm and F-Tree algorithm. We have also observed that the execution time of LargeItem has not a constant rule; as it will finish the execution when there is no further improvement of the total cost function. We have also observed that the execution time of LargeItem will grow at minimum support around 50%; and that growth comes as a result of the repetition of moving transaction *t* form one cluster to another, since it gives the same cost. On the other hand, the F-Tree algorithm grows in an exponential manner, and that is because the number clusters needs to merge increase proportionally with the minimum support.

We also perform sensitivity test of F-Tree on the order of input data using Mushroom. The result in Figure 5 is derived from the original data order. We test F-Tree with randomly ordered Mushroom data. The results are typical to the original ones. They show that F-Tree is not sensitive to the order of input data. However, our experiment results from randomly ordered Mushroom data show that LargeItem is more sensitive to data order.

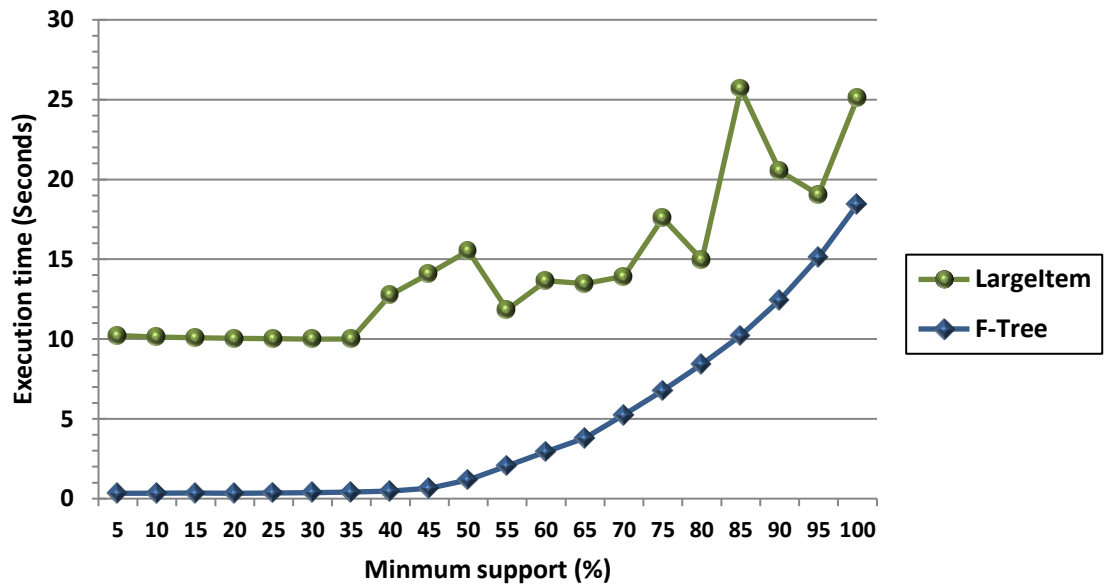

**Figure 5. Total Running Time of F-Tree and LargeItem on Mushroom dataset.**

For more details we examine the execution time of two algorithms in the allocation phase only. The result is shown in Figure 6. Here we observe that F-Tree in some manner is not highly depending on the minimum support in the allocation phase compared with the LargeItem algorithm. The execution time of the allocation phase of F-Tree seems to be constant. And the main reason behind this range of difference was the computation steps in LargeItem algorithm; that is used for assigning a new transaction *t* to the best existing clusters or creates a new cluster for it.

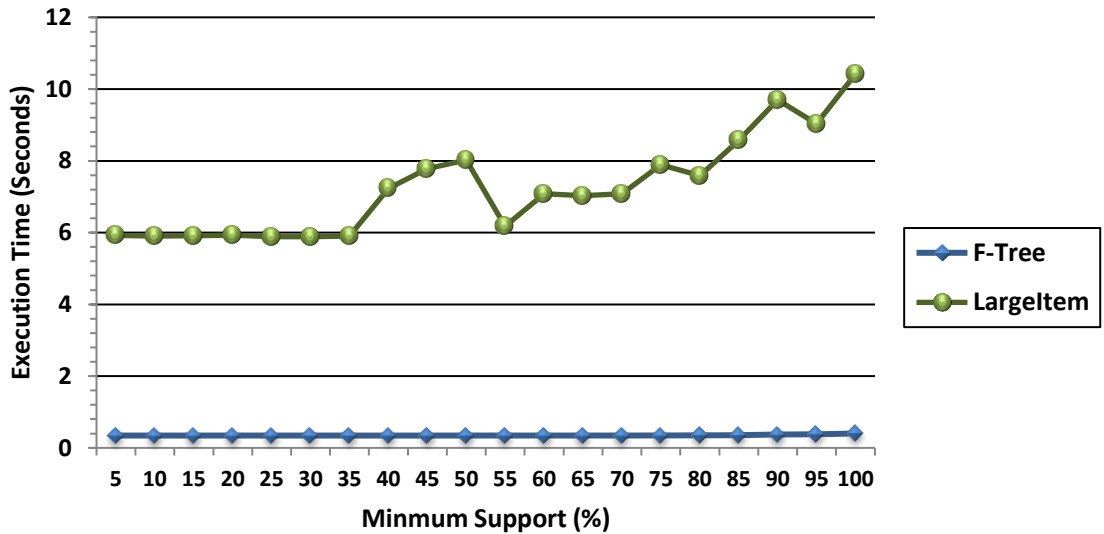

Figure 6. Allocation Phase Running Time of F-Tree and LargeItem on Mushroom dataset.

In Figure 7 , we detail the execution time of F-Tree in allocation phase. Here you can see more details about the running time of each step in the allocation phase. From this result we conclude that the first scanning of dataset to compute the frequencies is fast, while the in the second scan requires a little time to build the F-Tree structure. We also noted that extract clusters form F-Tree is very fast once the tree was built. But a little change resulting as the higher cut level $l$ of the F-Tree is increasing the traversing nodes. Generally, the execution time does not exceed half a second.

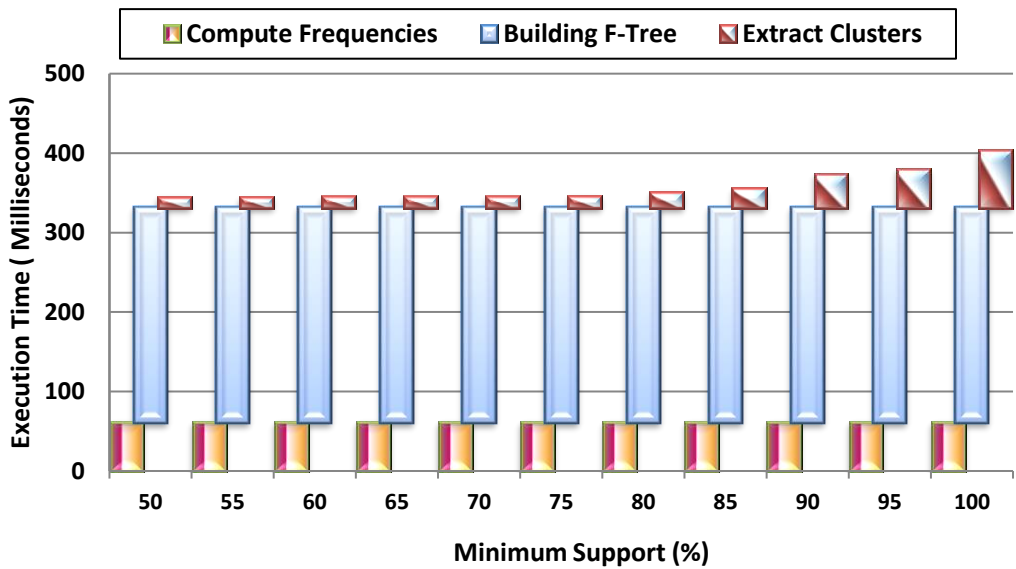

Figure 7. The Running Time Details of F-Tree allocation phase on Mushroom dataset.

For scalability experiment, we run F-Tree ($\alpha = 0.8$) and LargeItem ($w = 1$) with ($\theta = 0.2, 0.6, and\ 0.8$) on the 10%, 50% and 100% of the Mushroom respectively. The average of the running time is shown in Figure 8. We can see that the execution time of both F-Tree and LargeItem are linear to the dataset size. Form Figure 8 we can conclude that the F-Tree approach scales better with respect to dataset size.

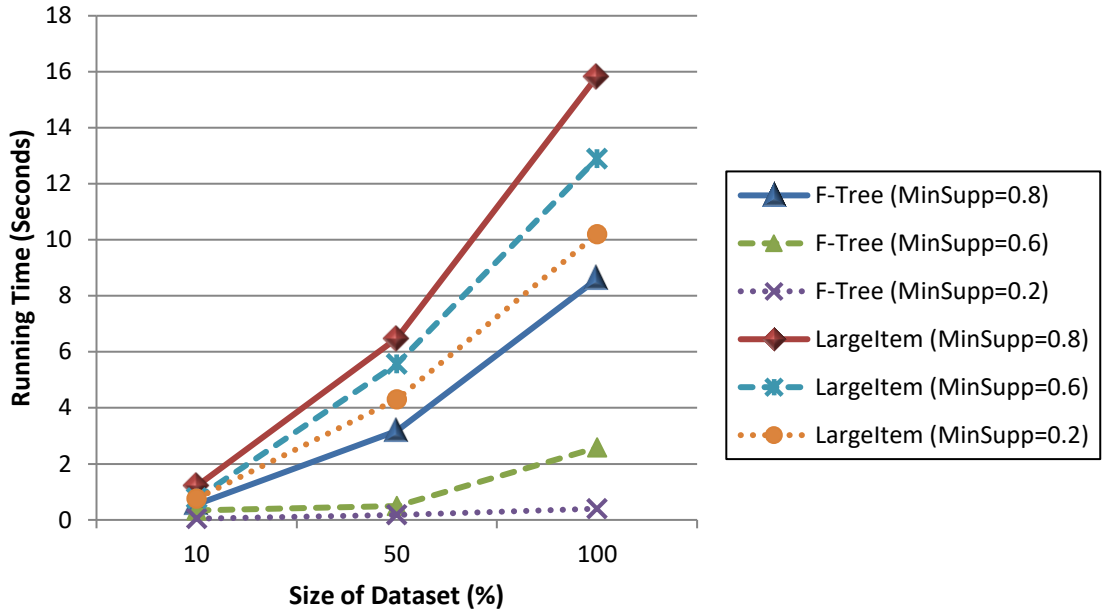

**Figure 8. Total Running Time of F-Tree and LargeItem on Mushroom with varying size.**

We compare the clustering quality of F-Tree on a mushroom dataset with those LargeItem [7], Seed [8], CLOP [2], and ROCK [4].

To evaluate the clustering quality we use the purity metric compute in [2] by summing up the larger one of number of "*edibles*" and the number of "*poisonous*" in every cluster. The number of clusters should be as few as possible, since a clustering with each transaction as a cluster will achieve a maximum purity [2].

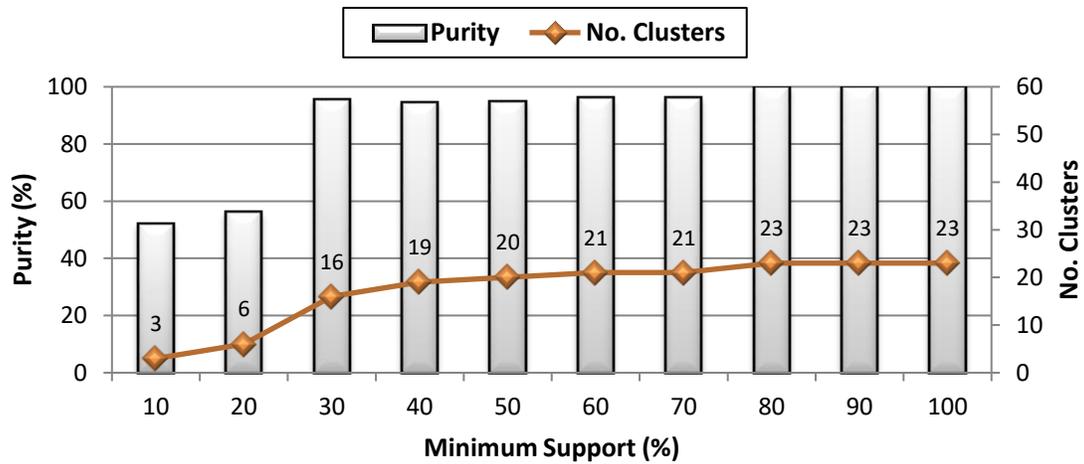

**Figure 9. The Number of Clusters vs. Clusters Purity result of F-Tree on Mushroom dataset.**

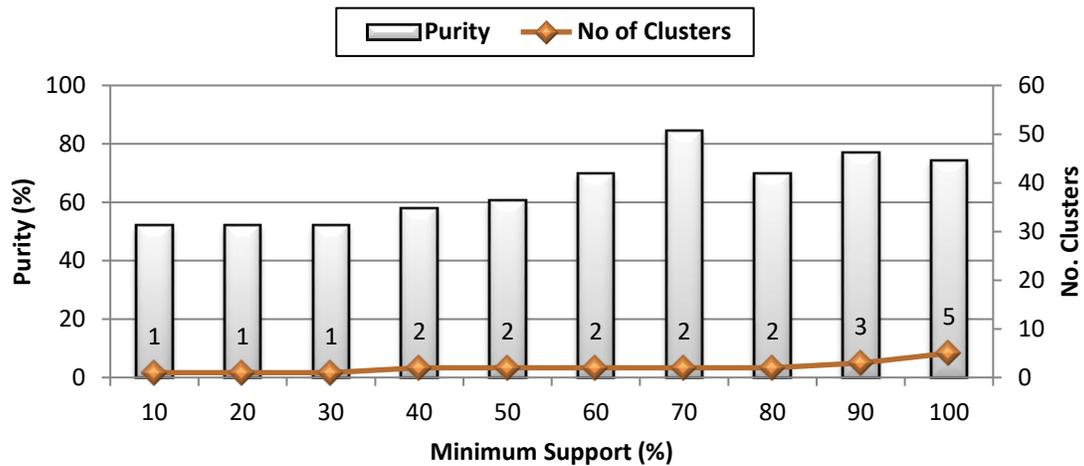

**Figure 10. The Number of Clusters vs. Clusters Purity result of LargeItem on Mushroom dataset.**

We try different values of minimum support $\theta$ from 10% to 100%, with $\alpha = 0.8$ the best value from the overlapping estimator. The results are shown in Figure 9.

When the minimum support $\theta = 80\%$, and here the cut level $l = 0.8 * 22 = 18$, the number of cluster is 23, and this is with a perfect classification (purity=100%). We also conclude that starting form minimum support $\theta = 30\%$, F-Tree result clusters purity above 90%.

The results of LargeItem, when a default weight of intra $w = 1$ was used, there is no good clustering was found with different $\theta$ from 10% to 100%. As shown in Figure 10. Our experiment shows that when we increased $w$ to make a larger intra more costly; we could find pure results at support 1 and $w = 10$, but with 58 clusters.

F-Tree results are quite close to results presented in the ROCK [4], where only result form [4] given is 21 clusters with only one impure cluster with purity=99.6%, by a minimum support $\theta = 80\%$. The results are also close to result presented in the CLOP [2], but CLOP gives a perfect classification with 30 clusters compared with F-Tree which gives only 23 pure clusters. CLOP given a 27 clusters but with one impure clusters with purity=99.6%. Table 10 shows the difference of those algorithms. Where r is a real number called repulsion, used to control the level of intra-cluster similarity in CLOP [2], $\theta$ is the minimum support, and $\alpha$ is the overlap threshold in F-Tree algorithm.

Table 10. The Purity and Number of Clusters results from different Algorithms.

| Clustering Algorithm | Criteria | Purity (%) | Number of Clusters |
|---|---|---|---|
| ROCK | $\theta = 80\%$ | 99.6 | 21 |
| CLOP | $r = 2.6$ | 100 | 30 |
| CLOP | $r = 3.1$ | 99.6 | 27 |
| F-Tree | $\theta = 80\%, \alpha = 0.8$ | 100 | 23 |
| F-Tree | $\theta = 70\%, \alpha = 0.8$ | 99.06 | 21 |
| F-Tree | $\theta = 40\%, \alpha = 0.8$ | 96.89 | 19 |
| F-Tree | $\theta = 30\%, \alpha = 0.8$ | 95.42 | 16 |

To observe the effect of the purity in the allocation phase in the Refinement phase; we have shown in Figure 11 and in Figure 12 the difference of clusters purity after the allocation phase and the refinement phase respectively in both F-Tree and Large item. We can see that F-Tree generate a high pure cluster in allocation phase starting form minimum support $\theta = 30\%$, while the LargeItem reaching a purity of 80% in at only minimum support $\theta = 70\%$.

After the refinement phase, we did not note a high change in the clusters purity in the LargeItem. But we observe that F-Tree have changed the purity of clusters in some cases as the result of merging clusters process which depend on $\alpha$ the overlap threshold value.

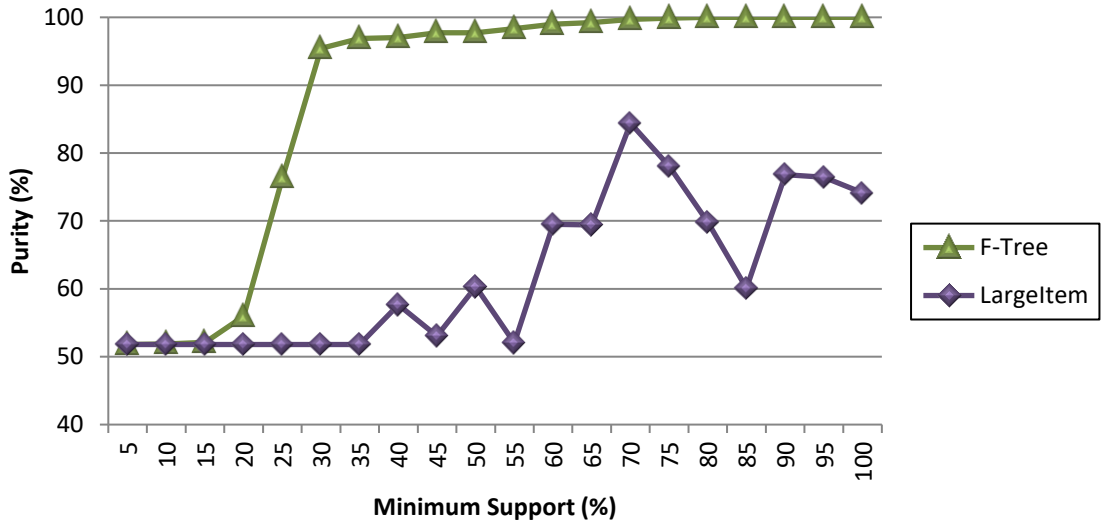

**Figure 11. The Results of Clusters Purity after the allocation Phase in F-Tree and LargeItem on Mushroom dataset**

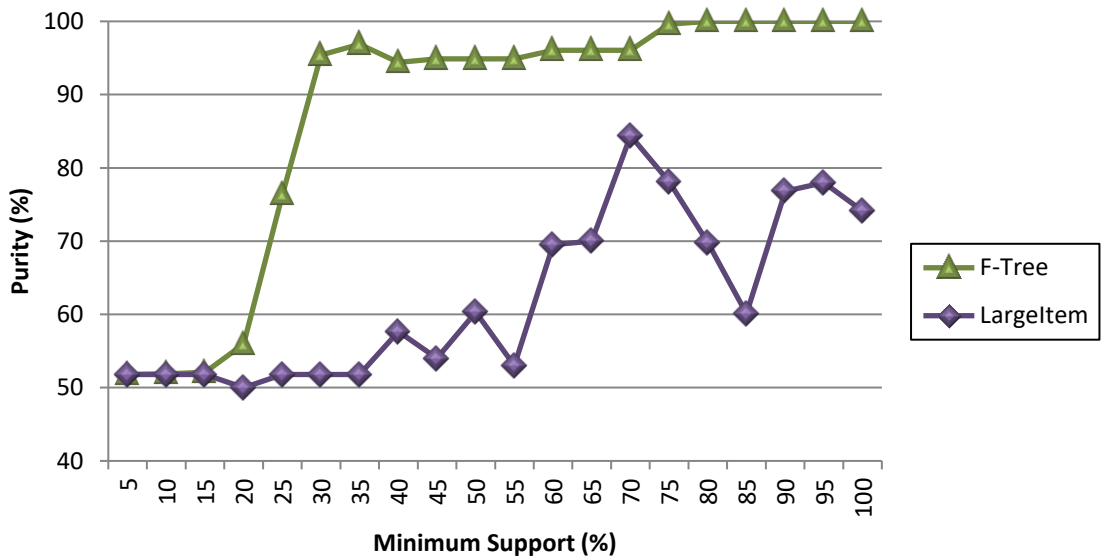

**Figure 12. The Results of Clusters Purity after the Refinement Phase in F-Tree and LargeItem on Mushroom dataset**

To illustrate the merging process; we count the number of clusters before and after the refinement phase (merging phase) along with the minimum support and showing the test result in Figure 13. As we illustrated when the minimum support $\theta = 100\%$, the number of clusters equal to the number of transactions. From Figure 13, we concluded that however the increasing in the minimum support increase the number of cluster generated. The number of clusters after merging did not increase in the same manner, which makes it lead to high purity of clusters.

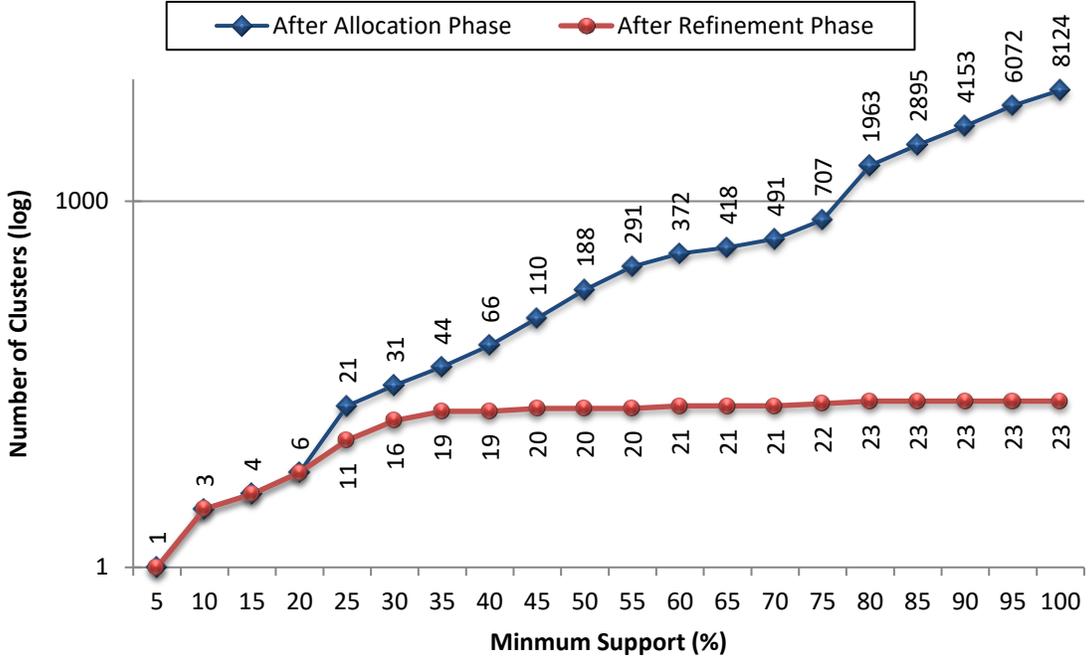

**Figure 13. Number of Clusters Result before and after the Refinement Phase of F-Tree on Mushroom dataset.**

### 5.3 Evaluation of clustering

In this section we report on cluster quality, as measured by the Root Mean Square Standard Deviation RMSSTD [5].

The RMSSTD is the square root of the variance of all the variables attribute of the clusters [3]. RMSSTD is measure the homogeneity of the clusters to identify the homogeneous groups; the lower RMSSTD value means the better clustering [13] [14]. The RMSSTD is given by equation (14), where $n_c$ is the number of clusters, $v$ the number of variables (data dimensionality), $n_j$ corresponds to the number of data values of $j$ dimension that belong to cluster $i$. also $\overline{x_k}$ is the mean of data values of $j$ dimension.

$$\boldsymbol{RMSSTD} = \frac{\sum_{\substack{i=1\ldots n_c \\ j=1\ldots v}} \sum_{k=1}^{n_{ij}}(x_k - \overline{x_k})^2}{\sum_{\substack{i=1\ldots n_c \\ j=1\ldots v}}(n_{ij} - 1)} \qquad (14)$$

We run experiments on the same datasets that have been used in, Mushroom and Zoo. Table 11 below shows that our clustering approaches in term of cluster quality. The F-Tree returned a lower RMSSTD values than LargeItem and Cluster Seeding. F-Tree gives the best quality across datasets. As a result of high purity of clusters which make the items of the clusters more homogeneous.

Table 11. RMSSTD results

| Dataset | No Clusters | No Transaction | F-Tree | | Cluster Seeding | | LargeItem | |
|---|---|---|---|---|---|---|---|---|
| | | | No Clusters | RMSSTD | No Clusters | RMSSTD | No Clusters | RMSSTD |
| Zoo | 7 | 101 | 9 | 5.18 | 7 | 21.2 | 7 | 24.9 |
| | | | 7 | 8.29 | | | | |
| Mushroom | 2 | 8124 | 23 | 20.01 | 8 | 28.5 | 6 | 30.0 |

# 6. CONCLUSION

We have presented F-Tree a fast, high quality and scalable algorithm for clustering transactional data. We have reported our experimental evaluation results real datasets. We show that compared to existing transactional data clustering methods, the F-Tree approach (the F-Tree clustering algorithm powered by the FSCO framework) can generate high quality clustering results in a fully automated manner with much higher efficiency for wider collections of transactional datasets. We have also proposed a new metric that controlling the similarity between clusters. The new metric depends on the overlap probability of cluster's items; which solve problem behind the distribution of items inside the clusters.

# 7. REFERENCES

[1] *Efficiently Clustering Transactional Data with weighed Coverage Density.* **H., Yan, K., Chen and L., Liu.** Virginia, USA : 'CIKM'06, ACM, Arlington, Nov.2006. Proceedings of the 15th ACM international conference on Information and knowledge management.

[2] *CLOPE: A Fast and Effective Clustering Algorithm for Transactional Data.* **Y., Yang, X., Guan and J., You.** Alberta, Canada : ACM SIGKDD '02, July 23-26, 2002. Proc. Of ACM SIGKDD Conference, Edmonton.

[3] *Data clustering: a review.* **A. K., Jain, M. N., Murty and P. J., Flynn.** 3, s.l. : ACM Comput. Surv., Sep. 1999, Vol. 31, pp. 264-323.

[4] *ROCK: A Robust Clustering Algorithm for Categorical Attributes.* **S., Guha, R., Rastogi and K., Shim.** Sydney, Australia : ICDE. IEEE Computer Society, March 23-26, 1999. In Proceedings of the 15th international Conference on Data Engineering, Washington, DC, 512. pp. 345-366.

[5] **S., Sharma.** *Applied multivariate techniques.* New York, NY, USA : Incorporation, John Wiley & Sons, 1996.

[6] *Combined use of association rules mining and clustering methods to find relevant links between binary rare attributes in a large data set.* **M., Plasse, et al.** 1, 2007, Science Direct, Computational Statistics & Data Analysis, Vol. 52, pp. 596-613.

[7] *Clustering transactions using large items.* **K., Wang, C., Xu and B., Liu.** New York, NY, USA : S. Gauch, Ed. CIKM '99. ACM, Nov. 1999. In Proceedings of the Eighth international Conference on information and Knowledge Management. pp. 483-490.

[8] *Transaction Clustering Using a Seeds Based Approach.* **Y.S., Koh and R., Pears.** Osaka, Japan : Springer, Heidelberg. Lecture Notes in Computer Science, May 20-23, 2008. Advances in Knowledge Discovery and Data Mining, 12th Pacific-Asia Conference, PAKDD 2008. Vol. 5021, pp. 916-922.

[9] *An Efficient Clustering Algorithm for Market Basket Data Based on Small Large Ratios.* **Yun., Ching-Huang, Chuang., Kun-Ta and Chen., Ming-Syan.** Washington, DC, USA : IEEE Computer Society, 2001. In 'COMPOSAC'01: Proceedings of the 25th International Computer Software and Applications Conference on Invigorating Software Development. pp. 505-510.

[10] *On the jaccard similarity test.* **G.I., Ivchenko and S.A., Honov.** 6, s.l. : Springer New York , March 1998, Journal of Mathematical Sciences, Vol. 88, pp. 789-794.

[11] *Rare Association Rule Mining via Transaction Clustering.* **Y.S., Koh and R., Pears.** Stamford Grand, Glenelg, Adelaide : s.n., Nov 2008. In Australasian Data Mining Conference: AusDM 2008.

[12] **D., Newman, et al.** UCI Repository of Machine Learning Databases. CA, USA : Department of Information and Computer Science, University of California, at Irvine, 1998.

[13] *Clustering Validity Checking Methods: Part2.* **M., Halkidi, Y., Batistakis and M., Vazirgiannis.** s.l. : SIGMOD. Record 31, 2002, pp. 19-27.

[14] **M., Halkidi, Y., Batistakis and M., Vazirgiannis.** Clustering Validity Checking Methods: Part1. s.l. : SIGMOD. Record 31, 2002, pp. 40-45.

[15] *Building E-Shop Using Incremental Association Rule Mining and Transaction Clustering.* **E., Anbalagan, E., Mohan and C., Puttamadappa.** 1, s.l. : Research India Publications, 2009, International Journal of Computational Intelligence Research, Vol. 5, pp. 11–23. ISSN 0973-1873.

[16] *A New Clustering Algorithm for Transaction Data via Caucus.* **J., Xu, et al.** Seoul, Korea : Springer, Heidelberg, April-May, 2003. In Advances in Knowledge Discovery and data Mining:7th Pacific-Asia Conference, 'Proceedings', PAKDD 2003. Vol. 2637, pp. 551-562.

[17] *A Fast Clustering Algorithm to Cluster Very Large Categorical Data Sets in Data mining.* **Z., Huang.** Australia : DMKD, 1997, Cooperative Research Center for Advanced Computational Systems, CSIRO Mathematical and Information sciences.